\documentclass{article}
\usepackage{graphicx}
\usepackage{color}

\usepackage{amsmath, amsthm, amsfonts}
\usepackage{amssymb}
\usepackage{mathrsfs}
\usepackage[a4paper,left=1.6cm,right=1.6cm,top=1.6cm,bottom=1.6cm]{geometry}
\usepackage{cite}
\usepackage{caption}

\title{Bending of Light from Reissner-Nordstr\"om-de Sitter-Monopole Black Hole }
\author{M. Haluk  Se\c{c}uk and \"Ozg\"ur Delice \\
  \small Physics Department, Faculty of Science and 
Letters, Marmara University, 34722  Istanbul, Turkey\\
  \small haluksecuk@marun.edu.tr\\
  \small ozgur.delice@marmara.edu.tr
}

\begin{document}
\maketitle

\abstract{We study light deflection from a Reissner-Nordstr\"om-de Sitter black hole in the gravitational monopole background. We first calculate the orbit equation and the contribution of the monopole and black hole parameters to the deflection angle up to second-order for the vanishing cosmological constant case using the Rindler-Ishak method.  We also obtain the contribution of the cosmological constant to light deflection in this geometry in the weak field limit using the same method.}  

\section{Introduction}\label{intro}
In a very recent observation, the researchers successfully obtained an image of the black hole centered at the galaxy M87\cite{event2019first}. The image shows a photon orbit around the black hole which made the observation possible in the first place since one can only observe the effects of the interaction of the surrounding matter distribution with a  black hole,  but not the black hole itself. This observation shows the importance of the interaction of light with the gravitational field of strong compact gravitational sources, which are very hard to detect with traditional astrophysical methods. This observation also shows the importance of understanding the deflection of light by different sources, such as the black holes, as in the first observational tests of Einstein's General Relativity in 1919 by the famous expedition lead by Dyson, Eddington and Davidson \cite{dyson1920determination} based on the weak gravitational lensing approximation calculated by  Einstein himself. After that observation,  gravitational lensing phenomena become one of the fundamental concepts in the general theory of relativity and astrophysics.

%----------------------------------------------------------------------------

 In this paper, we consider a global monopole swallowed by a massive, charged black hole, namely Reissner-Nordstr\"{o}m-monopole black hole in the de-Sitter spacetime. Global monopoles are one of the various types of topological defects which may be produced at the early universe in the symmetry-breaking phase transitions. Different types of topological defects, namely monopoles, cosmic strings, domain walls, or textures,  can be produced, depending on the types of broken symmetry. In theory, it is established that the monopoles are formed by a spontaneously broken O(3) symmetry to U(1)\cite{vilenkin2000cosmic,barriola1989gravitational}. One of the aspects of the gravitational monopoles is that the pure monopole spacetime is not asymptotically flat, having spacetime with a solid deficit angle. The gravitational effects of this exotic matter are different from ordinary matter. For example, the active gravitational mass of monopoles vanishes, hence exerting no force on the ordinary matter. However,  the topology of the spacetime is different from Euclidian one due to the solid angle deficit produced by the global monopole.  Therefore, the area of a sphere of proper radius $r$ is not $4 \pi r^2$, but rather $4 \pi r^2(1-8\pi\eta^2)$, where $\eta$ is the parameter determining the strength of the monopole. This feature of the gravitational monopole will play a role when we calculate the bending of light due to RN-monopole or RN-dS-monopole spacetimes. Due to this angle deficit, all light rays are deflected in the same manner \cite{barriola1989gravitational}. To better understand the deflection of light by a global monopole when it is not isolated, we consider the case where a monopole is swallowed by a Reissner-Nordstr\"om-dS black hole, which may imitate an astrophysical scenario where the monopole is not isolated but at the center of a spherical matter distribution such as a spherical galaxy or a black hole. 

%----------------------------------------------------------------------------
 
In this article, we calculate the deflection angle of light by a Reissner-Nordstr\"{o}m-monopole black hole. We first obtain and solve the orbit equation of this system by exploiting a perturbative method up to second order in the case of vanishing cosmological constant in Sect. 2. In Sect. 3, we calculate the deflection angle of the RN-monopole black hole using a well known and elegant method known as the  Rindler-Ishak method \cite{rindler2007contribution, ishak2010relevance}. Strong field gravitational  lensing for a  global monopole is discussed in \cite{Perlick:2003vg,Cheng:2010nd}. For a more general treatment, namely, to observe the effect of a possible positive cosmological constant, we again use the same method, obtain an analytical expression for the deflection angle of the RN-de Sitter-monopole solution and compare our results with the existing literature in Sect. 4. Furthermore, to determine the observational domain of the scattering procedure due to the limitations that come with Rindler-Ishak method, necessary constraint equations on the radial solution and the impact parameter of the incoming photons are also obtained in Sect. 4. As a result, we obtain analytical results for deflection angle up to second order in mass, charge, monopole term, and first-order in the cosmological constant in which some of them are new. We also compare the results we have obtained with the existing solutions in the literature, and we see that the results we have obtained a match with them perfectly. The paper finishes with a brief conclusion.

%----------------------------------------------------------------------------

\section{Calculation of Deflection Angle}

\subsection{The Line Element}

Global monopoles can be produced by a symmetry breaking phase transition where an $O(3)$ symmetry is broken to $U(1)$. Since we will only discuss the gravitational effects of these monopoles on light rays,  we refer the articles \cite{barriola1989gravitational,vilenkin2000cosmic} for details on global monopoles and also \cite{guendelman1991gravitational,Li:2002ku,Bertrand:2003yq} for the global monopoles in the presence of a cosmological constant.  Hence in this work, we concentrate on their gravitational field. The line element of the far-field of a global monopole swallowed by an RN-dS black hole was implicitly given in \cite{guendelman1991gravitational} as,
\begin{equation}\label{lineelement}
ds^2=-\frac{\Delta_r}{r^2}\,dt^2+\frac{r^2}{\Delta_r}\,dr^2+r^2\left(d\theta^2 + \sin^2\theta\,d\phi^2\right),
\end{equation}
with $\Delta_r$ is defined as,
\begin{equation}\label{Deltar}
\Delta_r=b^2r^2- 2Mr-\frac{\Lambda}{3}r^4+{Q^2},\quad 
b^2=(1- 8 \pi \eta^2)
\end{equation}
where $M$ and $ Q$ are the total mass and the total charge of the black hole, $\Lambda$ is the cosmological constant, and $\eta$ is the contribution of the global monopole, which are the physical parameters of this spacetime. Since the physically significant choice of monopole is $8\pi\eta^2<<1$, we have $0<b^2<1$. Notice that when one neglects all the parameters but gravitational monopole term, then by rescaling the $t \rightarrow t/b $ and $r\rightarrow b r$ variables one obtains the line element \cite{vilenkin2000cosmic,barriola1989gravitational},
\begin{equation}\label{assympmonmet}
ds^2=-dt^2+dr^2+(1- 8 \pi \eta^2)\,r^2\left(d\theta^2+\sin\theta^2d\phi^2\right).
\end{equation}
The metric \eqref{assympmonmet}, aside from describing the asymptotic behavior of the gravitational monopole outside the core, affirms that the pure gravitational monopole spacetime is not asymptotically flat and describes a spacetime with solid deficit angle, caused by the term $\eta$. The same spacetime (\ref{assympmonmet}) also describes a Letellier spacetime \cite{letelier1979clouds}, namely a  configuration where an ensemble of radially distributed straight cosmic strings intersecting at a common point which sometimes also called as the gravitational hedgehogs \cite{guendelman1991gravitational,delice2003gravitational}. Hence, the results we will obtain are also valid for this string-hedgehog configuration.

\subsection{Null Geodesics and The Orbit Equation}
The spacetime geometry we are considering is static and spherically symmetric. Hence we have a timelike Killing vector field $\chi=\partial_t$ generating staticity of the spacetime and also a rotational killing vector field  $\psi=\partial_\phi$ generating spherical symmetry. These symmetry generators lead to conserved quantities when the inner products of them with tangent vectors of the related geodesics are taken, which yield  $E=- g_{ab} \chi^a dx^b / d\lambda$ and $L=g_{ab} \psi^a dx^b / d\lambda$ where $dx^a/ d\lambda$  is the tangent vector of a given geodesics and $\lambda$ is an affine parameter of the geodesics.  Using these facts for the line element \eqref{lineelement}, we can find the  following equatorial ($\theta=\pi/2$) geodesic equations for a neutral lightlike  test particle,
\begin{eqnarray}
&&\frac{dt}{d\lambda}=-\frac{E\, r^2}{\Delta_r},\\
&&\frac{d\phi}{d\lambda}=\frac{L}{r^2},\label{phidot}\\
&&\frac{dr}{d\lambda}=\sqrt{E^2-L^2\frac{\Delta_r}{r^4}} .\label{radialgeo}
\end{eqnarray}
where %$\lambda$ is an affine parameter, 
$E$ and $L$ are specific energy and angular momentum of the null particle. Equation \eqref{radialgeo} can be put into the form
\begin{equation}\label{EnergyEquation}
\dot{r}^2=E^2-V_{eff},
\end{equation}
where overdot means the derivative with respect to the affine parameter, $\lambda$, and the effective potential is given by
\begin{equation}\label{EffectivePotential}
V_{eff}=L^2\frac{\Delta_r}{r^4}=\frac{L^2}{r^2} \left(b^2-\frac{2M}{r}-\frac{\Lambda\, r^2}{3}+\frac{Q^2}{r^2}\right). 
\end{equation}
Hence, radial geodesic  motion is only possible  for $E^2\ge V_{eff}$. 
Using the Eqs. \eqref{phidot} and \eqref{radialgeo}, one obtains an orbit equation as follows
\begin{equation}
\left(\frac{dr}{d\phi}\right)^2=\frac{r^4}{L^2}\left[E^2- L^2\frac{\Delta_r}{r^4} \right].
\end{equation}
With employing the usual inverse radial distance parameter $u=1/r$,
the orbit equation takes the following form
\begin{eqnarray}
\left(\frac{du}{d\phi}\right)^2=\frac{E^2}{L^2}- u^4\,\Delta_r(u) .
\end{eqnarray}
The explicit form of this equation is obtained if we replace the metric function $\Delta_r$,  which is
\begin{eqnarray}
\left(\frac{du}{d\phi}\right)^2=\frac{E^2}{L^2}-\left( b^2 u^2-2M u^3-\frac{\Lambda}{3}+Q^2u^4 \right) .
\label{explicitversion}
\end{eqnarray}
Note that the right-hand side of this equation involves corrections of the GR from monopole term, mass, the cosmological constant and electrical charge of the source to the flat space solution for photon motion. 
Taking the derivative of both sides with respect to the coordinate $\phi$, we obtain a modified Binet equation as follows:
\begin{equation}
\frac{d^2u}{d\phi^2}+u=(1-b^2)u+3M u^2-2Q^2 u^3.\label{modbinet}
\end{equation}
This equation does not involve  cosmological constant term, as first discovered for S-dS spacetime in \cite{Islam:1983rxp}. This result was interpreted that the cosmological constant term does not affect the null geodesics.   However,  bringing null geodesics equation in a form independent of cosmological constant is not enough to justify that the cosmological constant does not affect the null geodesics, since the solution of \eqref{modbinet} should also satisfy the governing Eq. \eqref{explicitversion} which clearly contains $\Lambda$. The effect of cosmological constant on null geodesics were investigated in detail in \cite{lebedev2013influence}. They showed that the null geodesics of S-dS spacetime can be made independent of $\Lambda$ depending on how to choose the boundary conditions of a given null geodesics. This can be easy to  see since in Eq. (\ref{explicitversion}) where  the term involving $\Lambda $ is just a constant, which can be clearly absorbed into the other constant $E^2/L^2$. Hence, it is possible to absorb the effect of cosmological constant, whose effect is negligible in  Eq. (\ref{explicitversion}) since the value of the observed cosmological constant is  extremely  small compared to the possible values of square of the energy per momentum of a photon $E^2/L^2$,   by redefining the integration constants as $E^2/L^2+\Lambda/3 \rightarrow  E'^2/L^2 $. Hence, Eq. \eqref{explicitversion} can be made independent of $\Lambda$ by shifting the energy per angular momentum of the photon, which does not change the classical orbit in a meaningful way. The idea that the cosmological constant does not affect the null geodesics in S-dS spacetime is a choice most of the works in the literature \cite{rindler2007contribution,sereno2008influence} follows. For a slightly different approach on this topic, see \cite{zhao2016gravitational}. For clarity, we first postpone  the effect of the cosmological constant to this phenomena to the Sect. 4 and restrict ourselves with a solution describing monopole swallowed by a Reissner-Nordstr\"om black hole by setting $\Lambda=0$ in the  Sect. 3.

%%%%%%%%%%%%%%%%%%%%%%%%%%%%%%%%%%%%%%%%%%%%%%%%%%%%%%%%%%%%%%%%%%%%%%%%%%%%%%%%%%%%%%%%

\subsection{The  Solution of the Orbit Equation}

In the previous subsection, we have found the orbit equation for null geodesics in Binet form with $u=1/r$ in Eq. (\ref{modbinet}). Now we will analyze the deflection of light rays from  RN-monopole black hole configuration. We will first employ a perturbative approach to find the orbit equation. Hence, we will use the following perturbation parameters
\begin{eqnarray}
\varepsilon=\frac{M}{R},\quad
\nu=\frac{Q^2}{R^2},\quad
\eta'=8\pi\eta^2=1-b^2,
\end{eqnarray}
which will help us to keep track of the order of the perturbation parameters. Note that here we apply the perturbation scheme relative to the flat spacetime, namely we will treat the monopole term $\eta'$ as another source of the curved spacetime similar to mass or charge of the black hole. Hence, here $R=L/E$ is the impact parameter of the flat background spacetime. Then the  resulting equation is,
\begin{equation}\label{Binetlight}
\frac{d^2u}{d\phi^2}+u=\eta' u+3\varepsilon R \, u^2- 2 \nu R^2 u^3.
\end{equation}
Now we exploit the usual perturbative technique up to second-order terms to determine a solution to Eq. (\ref{Binetlight}). Hence, we consider the following solution ansatz which contains  solutions up second-order in the perturbation parameters $\eta',\varepsilon,$ and $\nu$: 
\begin{eqnarray}\label{secondorderansatz}
u(\phi)&=&u_0(\phi)+ \eta'\, u_b(\phi)+\varepsilon\, u_m(\phi)+ \nu\,  u_q(\phi)+\eta'^2\, v_b(\phi)+\varepsilon^2\, v_m(\phi)\\
&&+\nu^2\, v_q(\phi)+\eta' \varepsilon\, z_{bm}(\phi)+\eta' \nu\, z_{bq}(\phi)+\varepsilon \nu\, z_{mq}(\phi).\nonumber
\end{eqnarray} 
By replacing (\ref{secondorderansatz}) into (\ref{Binetlight}), after some lengthy calculation process, we find the following solution, up to second-order in the perturbation parameters, as
\begin{eqnarray}
u(\phi)&=&\frac{\cos\phi}{R}+\eta' \frac{\cos (\phi )+\phi  \sin (\phi )}{2 R} - \varepsilon \frac{\cos (2 \phi ) -3}{2 R}+ \nu \frac{\cos (3 \phi )  -12 \phi  \sin (\phi )-9 \cos (\phi )}{16 R}\nonumber \\
&&+\eta'^2 \frac{\left(3 - \phi ^2\right) \cos (\phi )+ 3 \phi  \sin (\phi )}{8 R} + \varepsilon^2 \frac{  60 \phi  \sin (\phi )+ 37 \cos (\phi )+3 \cos (3 \phi )}{16 R}\nonumber \\
&& +\nu^2 \frac{ \left(271-72 \phi ^2\right) \cos (\phi )+384 \phi  \sin (\phi )-36 \phi  \sin (3 \phi )-48 \cos (3 \phi )+\cos (5 \phi )}{256 R}\nonumber \\
&&-\eta'\varepsilon \frac{\phi  \sin (2 \phi )+2 \cos (2 \phi )-6}{2 R}+\eta'\nu \frac{ 3 \left(4 \phi ^2 - 15\right) \cos (\phi )+3 \phi  [\sin (3 \phi )-19 \sin (\phi )]+5 \cos (3 \phi )}{32 R} \nonumber\\
&&- \varepsilon \nu \frac{87-40 \cos (2 \phi )-12 \phi  \sin (2 \phi )+\cos (4 \phi )}{16 R}. \label{secondordersolution}
\end{eqnarray}
The above solution is obtained in a way that the path is symmetric for $\phi\rightarrow -\phi$. Hence, we have discarded $\sin{\phi}$ terms in the solution. Note also that when solving the orbit Eq. (\ref{Binetlight}) up to second-order in the parameters,  the obtained solutions may involve some arbitrary functions of the zeroth- and first-order solutions of this equation as well. However, to cure this problem,  one can determine these arbitrary functions by using the governing equation of (\ref{Binetlight}), namely Eq. (\ref{explicitversion}) (for $\Lambda=0$ of course). Hence, solution (\ref{secondordersolution}) is obtained in a way that it solves both of the orbit Eqs. (\ref{Binetlight}) and (\ref{explicitversion}), simultaneously. The solution we have obtained  \eqref{secondordersolution} reduces to the corresponding solution for RN case if the monopole term $\eta'$ is vanishing, given in ref. \cite{gergely2009second}. 

\section{Light Deflection From RN-Monopole Spacetime}

In this part, we will discuss the deflection angle of the spacetime geometry we consider. Consider a photon which comes from far away at a distant past ($u=0,\phi=-\pi/2-\delta\phi_a/2$) deflected by the black hole and travels towards far away  at distant future ($u=0, \phi=\pi/2+\delta\phi_a/2$) where $\delta\phi_a$ is an angle whose role will be discussed later in the paper.  Since solution (\ref{secondordersolution}) is symmetric under the transformation $\phi \rightarrow -\phi$, we can calculate $\delta\phi_a$  by feeding $\phi\rightarrow \pi/2+\delta\phi_a/2$ into (\ref{secondordersolution}) and  taking the limit  $\delta\phi_a\rightarrow 0$. Using this fact, we find the following result
\begin{eqnarray}\label{deltaphia}
\delta\phi_a=\frac{\pi}{2}\eta'+4\varepsilon-\frac{3\pi}{4}\nu+\frac{3\pi}{8}\eta'^2+\frac{15\pi}{4}\varepsilon^2+\frac{105\pi}{64}\nu^2+8\eta'\varepsilon-\frac{15\pi}{8} \eta'\nu-16 \varepsilon \nu.
\end{eqnarray}
Now {\emph{ if the spacetime we consider was asymptotically flat, then $\delta\phi_a$ would be the deflection angle.}} However, RN-monopole spacetime is {\bf{not asymptotically flat}} and $\delta\phi_a$ is {\bf not} the deflection angle of the RN-monopole black hole spacetime. In other words, Eq. (\ref{deltaphia}) does not fully reflect the effect of the monopole term on the bending of light. However as we will show, it is still related to the deflection angle. 

 We can still obtain the full contribution of the monopole to light bending by using another method. The most straightforward way is to use the Rindler-Ishak method \cite{rindler2007contribution},  which was first presented to find the effect of the cosmological constant on the bending of light where, similar to our case,  the spacetime is not asymptotically flat. 
 \begin{figure}[htp]
 	\begin{centering}
 		\includegraphics[draft=false,width=7cm]{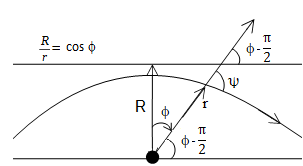}
 		\caption{The geometric setting of the Rindler-Ishak method.
 		Since we have a symmetrical solution in $\phi$, we have rotated
 		the framework by $-\frac{\pi}{2}$, and as a result, the half
 		bending angle is now given by  $\epsilon=\psi-\phi+\frac{\pi}{2}$
 		and for $\phi=\frac{\pi}{2}$ we recover $\epsilon=\psi$.}
 		\label{f1}
 	\end{centering}
 \end{figure}

Although the effect of the cosmological constant (more generally the spacetimes that are not asymptotically flat) on the light bending is a controversial  topic, the method developed by Rindler and Ishak seems to be  \cite{rindler2007contribution} the most elegant way to determine such contribution. Hence, to observe the complete effect of the monopole to the light deflection we follow the Rindler and Ishak method, and  we refer Fig. \ref{f1} for the explanation of the angles we will discuss. Their method  considers a special treatment where the source and observer are taken to be static\cite{lebedev2013influence}. Since the form of the line element is in the same form with \cite{rindler2007contribution}, we do not repeat their calculational steps  in this paper for clarity and directly apply the metric function $\Delta_r$ to their formula.    For small angles, we adopt Eq. (16) of \cite{rindler2007contribution}  which  reads the bending angle for the line element (\ref{lineelement}) as,
\begin{align}
\tan\psi = \sqrt{\frac{\Delta_r}{r^2}}\, r \left| \frac{dr}{d\phi} \right|^{-1}
\rightarrow\quad
\psi \approx \left[\sqrt{\frac{\Delta_r}{r^2}}\, r \left| \frac{dr}{d\phi} \right|^{-1}\right]_{\phi=\frac{\pi}{2}}.
\label{apprxangle}
\end{align}
Then,  we calculate,
\begin{eqnarray} \label{rpiover2}
\left[ \frac{1}{r} \right]_{\phi=\frac{\pi}{2}} &=& \frac{2 \, \varepsilon}{R}
\left(    
		1 + 2 \eta' - 4 \nu
	- \frac{ 15 \pi  \eta'\nu }{32 \varepsilon }
	+ \frac{ \pi  \eta' }{ 8 \varepsilon }
	- \frac{ 3 \pi \nu }{ 16 \varepsilon }
	+ \frac{ 3 \pi  \eta'^2}{ 32 \varepsilon }
	+ \frac{ 15 \pi  \varepsilon }{16}
	+ \frac{ 105 \pi  \nu ^2}{256 \varepsilon }
\right),
\\
\left| \frac{dr}{d\phi} \right|_{\phi=\frac{\pi}{2}} &=&\frac{R}{\,\, 4\,\varepsilon^2}
\left(    
		1 + 2 \eta' - 4 \nu
	- \frac{ 15 \pi  \eta'\nu }{32 \varepsilon }
	+ \frac{ \pi  \eta' }{ 8 \varepsilon }
	- \frac{ 3 \pi \nu }{ 16 \varepsilon }
	+ \frac{ 3 \pi  \eta'^2}{ 32 \varepsilon }
	+ \frac{ 15 \pi  \varepsilon }{16}
	+ \frac{ 105 \pi  \nu ^2}{256 \varepsilon }
\right)^{-2}
\nonumber \\ & \times &
\left(   
		 1- \frac{\,\,  \pi^2 }{32}\, \eta'^2 
       - 2 \, \varepsilon^2 
       - \frac{\,\, 9 \pi^2 }{128}\, \nu^2
       - \frac{\pi}{2}\, \eta' \varepsilon 
       + \frac{\,\, 3 \pi^2 }{32}\, \eta' \nu 
       + \frac{3 \pi}{4}\, \varepsilon \nu       
\right)\label{drpiover2}.
\end{eqnarray}
Inserting these results and also   the metric function $\Delta_r$ given in Eq. (\ref{Deltar}) to Eq. \eqref{apprxangle}, and expanding them in the weak field approximation we obtain the  half bending angle as
\begin{align}
\psi &\approx 
\left(
	1 - \frac{\eta' }{2} - \frac{M}{r} + \frac{Q^2}{2 r^2} 
	- \frac{\eta'^2}{8} - \frac{M^2}{2 r^2} - \frac{Q^4}{8 r^4} 
	- \frac{\eta' M}{2r} + \frac{\eta' Q^2}{4 r^2} + \frac{M Q^2}{2 r^3} 
\right) 
	r \left| \frac{dr}{d\phi} \right|^{-1},
\\ &   
= \frac{\delta \phi_a}{2} - 
\left\lbrace
	\frac{\pi}{8}\eta'^2 + \frac{\eta' M}{R} - \frac{3 \pi \eta' Q^2}{16 R^2}
\right\rbrace
	+ O(\eta^3,\varepsilon^3,\nu^3).
\end{align}
Here when evaluating the above expression, we have evaluated $r(\phi)$ at $\phi=\pi/2$ as given in Eq. \eqref{rpiover2} and expanded the metric function in \eqref{apprxangle} to the second order. Since we have chosen the parameter $\eta'$ as another source of the spacetime, in the first order, its effects are already included in the deflection angle $\delta\phi_a$. Hence, in this setting, the effect of the monopole due to the solid angular deficit reveals itself in the second-order terms involving $\eta'$ as we have expected.  Hence, the total deflection angle  $2\psi$  up to second-order in terms of parameters of the black hole, monopole and the impact parameter $R$ are found as
\begin{eqnarray}\label{deltaphim}
\delta\phi_M = 2 \psi &=& 
4 \pi^2 \eta^2 + \frac{ 4M }{ R } - \frac{ 3\pi Q^2 }{ 4R^2 } 
+ 8 \pi^3 \eta^4 + \frac{ 15\pi M^2 }{ 4 R^2 } + \frac{ 105\pi Q^4 }{ 64 R^4 }
\nonumber \\
&& + \frac{ 48 \pi\eta^2 M }{ R }-\frac{ 12\pi^2 \eta^2 Q^2 }{ R^2 } - \frac{ 16 M Q^2 }{ R^3 }.\label{secondordersolution1}
\end{eqnarray}
On the other hand, we can also calculate the minimum distance, $r_{min}$, that the light can approach the central deflecting object to express the deflection angle. Setting $\phi=0$ in (\ref{secondordersolution}) yields $u=1/r_{min}$, with the result
\begin{eqnarray}\label{r:min}
\frac{1}{r_{min}}&=&\frac{1}{R}\left(1+\frac{\,\eta'}{2}+\frac{M}{R}-\frac{Q^2}{2 R^2} +\frac{\,\,3 \eta'^2}{8}+\frac{5 M^2}{2 R^2}
\nonumber \right.\\
&& \left.+\frac{7 Q^4}{8 R^4}+\frac{2\eta' M}{R}-\frac{5 \eta'Q^2}{4 R^2}-\frac{3MQ^2}{R^3}\right).
\end{eqnarray}
Now we need to invert this equation for $r_{min}$, then put back into (\ref{secondordersolution1}). It turns out that  we will only need  the first-order correction terms for the parameters, namely we just need the following expression
\begin{equation}
\frac{1}{R}\simeq\frac{1}{r_{min}}\left(1-\frac{\,\eta'}{2}-\frac{M}{r_{min}}+\frac{Q^2}{2r_{min}^2} \right).
\end{equation}
Replacing this into (\ref{secondordersolution1}), we find the deflection angle, in terms of minimum distance, in the order $O(\eta'^2, \varepsilon^2, \nu^2)$, as follows:
\begin{eqnarray}
\delta\phi_M&=& 4\pi^2\eta^2+\frac{4M}{r_{min}}-\frac{3\pi Q^2}{\,4{r_{min}^2}}+8\pi^3\eta^4+\frac{(15\pi-16) M^2}{4r_{min}^2}+\frac{57\pi Q^4}{\,64 r_{min}^4}\nonumber \\
&&+\frac{32 \pi \eta^2 M}{r_{min}}-\frac{6 \pi \eta^2 Q^2}{ r_{min}^2}+\frac{(3\pi-28)MQ^2}{2r_{min}^3}.\label{lightdeflection}
\end{eqnarray}

Hence, we have found the deflection angle up to second-order terms in our perturbation parameters, $\eta',\varepsilon$ and $\nu$. When we compare this result with the existing solutions in the literature, we see that it agrees with most of the previous results found before. The first term and fourth terms are the effects of the monopole, i.e., the angle deficit, in the photon motion, and are compatible with the result given in \cite{harari1990repulsive}. The terms first order in the mass of the black hole is the famous result derived by Einstein himself, see also \cite{Rindlerbook}, and the second-order term in mass is the PPN result, \cite{epstein1980post, fischbach1980second, richter1982second} and also presented in \cite{edery2006second}. The first-order effects of the electrical charge are presented in \cite{briet2008determining} in brane-world models where a black hole living in a brane embedded in five-dimensional bulk resembles the form of an RN black hole of GR \cite{Dadhich:2000am}. The second-order corrections of the charge to the light deflection are also presented in \cite{gergely2009second} again for a brane-world scenario.
The deflection of light rays is also calculated recently in \cite{jusufi2017light}  for a solution describing rotating uncharged black hole with a monopole term presented in \cite{teixeira2001gravitational}, using the Gauss-Bonnet method introduced in \cite{Gibbons:2008rj}. The result they have obtained, i.e., the term proportional to $\eta^2 M$ has different numerical factor than our result in Eq. (\ref{secondordersolution1}). The reason of this ambiguity comes from the fact that we use Schwarzschild type coordinates in this work, but in \cite{jusufi2017light}, the authors employ coordinates rescaled by monopole term as $r\rightarrow \sqrt{(1-8\pi\eta^2)}r$ and $M\rightarrow M/(1-8\pi\eta^2)^{3/2}$ since they follow the convention of \cite{teixeira2001gravitational}.  However, we have checked that when their result is transformed into the coordinates in our paper, the numerical factors of $\eta^2 M$ term agree. Note also that, as they have stressed in their paper, due to the limitations of the method they use,  some of the second-order terms we have obtained in Eq. (\ref{deltaphim}), namely, the terms proportional to $M^2$ and $\eta^4$,  are not present in their results. This is expected because, as they have said in their work, the method they have used in \cite{jusufi2017light} only works in the linear order. Hence our results, unlike \cite{jusufi2017light}, given in Eq. (\ref{deltaphim}), properly generalize the deflection angle up to the second-order terms due to the monopole term.  
In astrophysical setting, the second-order terms (i.e., $O(\nu^2)$) in electrical charge (terms in the order of $Q^4$) may be irrelevant, but in brane-world scenarios, their effects may be too big to ignore. Hence, we have kept those terms for a possible application to such scenarios. We have seen that the effect of the monopole term is to increase the deflection angle and it acts as a magnifying lens similar to mass and contrary to the charge of the black hole.

%%%%%%%%%%%%%%%%%%%%%%%%%%%%%%%%%%%%%%%%%%%%%%%%%%%%%%%%%%%%%%%%%%%%%%%%%%%%%%%%%%%%%%%%

\section{Light Deflection From RN-dS-Monopole Spacetime}

In the previous section, we have ignored the effect of the cosmological constant on the deflection of light. As we have said in the previous section, the effect of the cosmological constant on the light bending is a controversial topic; the method developed by Rindler and Ishak seems to be  \cite{rindler2007contribution} the most elegant way to determine such contribution. The presence of a cosmological constant dramatically changes both the asymptotical structure of the spacetime and also the horizon structure. Since the roots of the equation $\Delta_r=0$ determine the horizons and since $\Delta_r$ given in Eq. (\ref{Deltar}) is fourth-order in $r$,  Vieta's formulas show that we have three real, positive, and one negative root. The real positive roots of this equation determine the black hole event ($r_+$), Cauchy ($r_-$), and the cosmological ($r_c$) horizons of this spacetime, respectively. The relevance of this discussion to the deflection of light is that the light rays deflected by the source cannot come from radial infinity outside the cosmological horizon. We will discuss the implications of this fact at the end of this section. Now, let us derive the deflection angle of the light rays due to RN-dS-monopole spacetime. 

 In this part, we will calculate the contribution of the positive cosmological constant on the deflection angle by considering the case $\Lambda\neq 0$ in (\ref{lineelement}). Since we have presented the necessary formulas to calculate the deflection angle in the previous section, we just present the necessary steps to obtain the results. Note that by re-defining the impact parameter as $1/R^2=E^2/L^2+\Lambda/3$ in Eq. (\ref{explicitversion}), both the governing and Binet equations take the same form with $\Lambda=0$ case. Hence, we can use the results of Sect. 2.3, and the orbit equation will be the same form for $\Lambda\neq 0$. We mainly need Eq. (\ref{apprxangle}) for the metric function $\Delta_r$ given in (\ref{Deltar}) which results
 
\begin{align}\label{Lambdacontrpsi1}
\psi& \approx \left[\sqrt{\frac{\Delta_r}{r^2}}\, r \left| \frac{dr}{d\phi} \right|^{-1}\right]_{\phi=\frac{\pi}{2}}= \left[\sqrt{1-\eta' -\frac{2M}{r}-\frac{\Lambda r^2}{3}+\frac{Q^2}{r^2} }   \, r \left| \frac{dr}{d\phi} \right|^{-1}\right]_{\phi=\frac{\pi}{2}}.
\end{align}
In this expression, $r=r(\phi)$ should be evaluated at $\phi=\pi/2$. To find the proper result, it turns out that we need to expand $\Delta_r$ given in Eq. (\ref{apprxangle}) and written explicitly in Eq. (\ref{Lambdacontrpsi1}),  up to second order in parameters as we have done in the RN-monopole case. Doing this in  Eq. (\ref{Lambdacontrpsi1}), we find that the extra terms containing the contribution of the cosmological constant to the light deflection half-angle have the following expression,  in addition to the RN-monopole contribution calculated in the previous section, as follows;
\begin{align}\label{Lambdacontrpsi}      
2\psi
\approx &\,	\delta\phi_M - \frac{\Lambda}{6}
 \left[
		\left(1 + \frac{\eta'}{2} \right) r^2
		+ M r - \frac{Q^2}{2}
\right] 
	 \left| \frac{dr}{d\phi} \right|^{-1}.
\end{align}
Here, the $\delta\phi_M$ term denotes all of the contributions of the mass, charge and monopole term in the light deflection angle up to second-order given in Eq. \eqref{deltaphim}. The second term in Eq. \eqref{Lambdacontrpsi} will give us the contribution of cosmological constant on the deflection angle when the terms  $r(\phi)$ and $dr/d\phi$ are  evaluated at $\phi=\pi/2$. The explicit forms of these terms are given in Eqs. (\ref{rpiover2},\ref{drpiover2}). This is the main result giving the contribution of $\Lambda$ to the light deflection.  The  total deflection angle can be formally written as 
\begin{equation}
2\psi = \delta\phi_M + \delta\phi_\Lambda,
\end{equation}
where  $\delta\phi_M$ is the bending angle due to mass, charge and monopole term given in Eq. (\ref{deltaphim}),  whereas the contribution from the cosmological constant, $\delta\phi_\Lambda$,  after some long calculations by replacing all terms in Eq. (\ref{Lambdacontrpsi}), can be written explicitly up to second order in $\Lambda, M, Q^2,\eta'=8\pi \eta^2$ in the following convenient form as,
\begin{equation}\label{contC}
\delta\phi_\Lambda = \Lambda 
\left( 
		L_0 + L_1 R + L_2 R^2 + L_3 R^3 + L_4 R^4 
\right).
\end{equation}

The explicit forms of the terms $L_0,\ldots, L_4$ are listed below where we have only kept up to first-order correction terms involving the parameters $M, Q^2$ and $\eta'$ since they are multiplied by $\Lambda$ in Eq. (\ref{contC}), which makes these terms up to second order in the parameters $M, Q^2,\eta',$ and $\Lambda$. Namely, we have 
\begin{align}
L_0 &=   
\frac{9 \pi^3 Q^6}{8192 M^4}+\left(\frac{163
   \pi }{512}-\frac{135 \pi ^3}{8192}\right)\frac{Q^4}{M^2}
   +\left(\frac{675 \pi ^3}{8192}-\frac{19 \pi
   }{16}\right)Q^2
\nonumber  \\ &    
   +\left[\frac{603 \pi ^3
   Q^6}{32768 M^4}-\left(\frac{1477 \pi
   }{1024}+\frac{4257 \pi ^3}{32768}\right)
   \frac{Q^4}{M^2}\right]\eta'   
   +\left(\frac{7965 \pi ^3}{131072}+\frac{45 \pi
   ^5}{131072}\right) \frac{Q^6}{M^4}\eta'^2,
\nonumber \\
L_1 &= \left(\frac{75 \pi^2}{512}-\frac{2}{3}\right)M+\frac{3 \pi ^2 Q^4}{512
   M^3}-\left(\frac{2}{3}+\frac{15 \pi ^2}{256}\right)
   \frac{Q^2}{M}+ \left[-\frac{163 \pi ^2 Q^4}{2048
   M^3}+\left(\frac{153 \pi ^2}{256}-3\right)
   \frac{Q^2}{M}\right]\eta'
\nonumber \\ &
   - \left(\frac{4709 \pi
   ^2}{8192}+\frac{9 \pi ^4}{8192}\right)
   \frac{Q^4}{M^4}\eta'^2,
\nonumber \\
L_2 &= \frac{5 \pi }{32} -\frac{\pi  Q^2}{32 M^2}
   + \left[-\frac{9
   \pi ^3 Q^4}{4096 M^4}+\left(\frac{45 \pi
   ^3}{2048}-\frac{37 \pi }{96}\right) \frac{Q^2}{M^2}-\frac{225
   \pi ^3}{4096}+\frac{127 \pi }{192}\right]\eta'
\nonumber \\ &
   + \left[-\frac{555\pi^3 Q^4}{32768 M^4}
   +\left(\frac{289 \pi }{384}+\frac{627
   \pi ^3}{8192}\right) \frac{Q^2}{M^2}\right]\eta'^2,
\nonumber \\ 
L_3 &= -\frac{1}{6 M} +\left(-\frac{\pi ^2 Q^2}{128 M^2}+\frac{5 \pi^2}{128}
   + \frac{1}{4}\right)\frac{\eta'}{M}
   + \left(\frac{25 \pi ^2 Q^2}{512 M^2}
   -\frac{87 \pi ^2}{512}+\frac{5}{6}\right)\frac{\eta'^2 }{M}. 
\nonumber \\
L_4 &= \frac{\pi}{48}\frac{ \eta'}{M^2}
   + \left(\frac{3 \pi^3 Q^2}{2048 M^2}
   -\frac{15 \pi ^3}{2048}+\frac{7 \pi}{64}\right)\frac{ \eta'^2 }{M^2}.
\nonumber
\end{align}
The terms involving higher powers of $M$ or $Q^2$ have resulted from a lengthy process of multiplication and expansion of the terms in (\ref{Lambdacontrpsi}).
Now let us analyze these terms in detail. The original contribution of the cosmological constant to the light deflection calculated by Rindler and Ishak in   \cite{ishak2010relevance,rindler2007contribution} is encoded in the first term in $L_3$.
Our results are also compatible with \cite{Arakida:2011ty} which also considered deflection of light in S-dS geometry, namely the first terms in $L_2$ and $L_3$ agree with their results given in Eq. (C2) in their work.  The first-order contribution of the electrical charge is calculated in \cite{article}. If we expand the result given by Eq. (15) of the \cite{article}  our results agree for charge coupling to the cosmological constant as given in the second term in $L_2$. The other charge contributions in our expressions are coming from second-order expansion terms which is not present in \cite{article} since in that work only the first-order expansion is studied. The only work which considers second-order expansion in the parameters in the presence of a cosmological constant which is comparable to our work is done by  \cite{Sultana2013}. When we compare our results with \cite{Sultana2013} we see that there is a strong resemblance except for a few terms. The reason of this mismatch is the fact that the orbit solution in Eq. (10) of \cite{Sultana2013} has a missing term proportional to $37 \sin \phi$  in the part proportional to second order in black hole mass. 
This contribution can be obtained by considering the following fact: Solution \eqref{secondordersolution} to Binet equation \eqref{Binetlight} \textit{should also} satisfy the orbit Eq. \eqref{explicitversion} up to the order of the perturbation, since Eq. \eqref{Binetlight} is obtained from Eq. \eqref{explicitversion}. This term is also absent in some other prior works \cite{Ishak:2007ea,ishak2010relevance}. Actually, the correct form of the perturbation solution second order in mass was already presented in \cite{gergely2009second}, which uses a different form for the solution, namely solution involving even functions,  as we have also followed in this work in Eq. (\ref{secondordersolution}). The correct form for the second-order orbit solution in mass is also presented later in \cite{Arakida:2011ty}, in the same form for the solutions used in  \cite{rindler2007contribution,Sultana2013}. Hence, our results correct the fifth and sixth terms of  Eq. (17)  of \cite{Sultana2013} with the first two terms of $L_1$  in our paper. The other relevant terms of  \cite{Sultana2013} agree with our results, namely the first terms of $L_2$ and $L_3$ are the same as the fourth and seventh terms of Eq. (17). These are the relevant works following similar methods with our work to find the contribution of the cosmological constant to the light bending. 

 For an astronomically relevant source satisfying $M\gg Q$ or for the very small or vanishing electrical charge,  the above terms simplify to 
 \begin{align}
 L_0 &= 0,
 \nonumber \\
 L_1 &= M \left( \frac{75 \pi
 	^2}{512}-\frac{2}{3}\right),
 \nonumber \\
 L_2 &= \frac{5 \pi }{32} 
 +\eta' \left[\frac{127 \pi }{192}-\frac{225
 	\pi ^3}{4096}\right]
 \nonumber \\
 L_3 &= -\frac{1}{6 M} +\frac{\eta'}{M} 
 \left(\frac{5 \pi^2}{128}
 +\frac{1}{4}\right)+ \frac{\eta'^2 }{M}\left(\frac{5}{6}
 -\frac{87 \pi ^2}{512}\right). 
 \nonumber \\
 L_4 &= \frac{1}{M^2} \left[\frac{\pi  \eta' }{48}
 +\eta'^2 \left(\frac{7 \pi}{64}
 -\frac{15 \pi ^3}{2048}\right)
 \right].
 \nonumber
 \end{align}
 These results clearly show the contribution of the monopole term to the deflection angle of the cosmological constant. As a result, we have generalized the previous works done by the cited authors via the inclusion of the monopole term and its couplings to the black hole parameters and cosmological constant up to second-order.

%%%%%%%%%%%%%%%%%%%%%%%%%%%%%%%%%%%%%%%%%%%%%%%%%%%%%%%%%%%%%%%%%%%%%%%%%%%%%%%%%%%%%%%%

\subsection{Domain of Validity} 
 One last but important note we need to make is the domain of validity of the result we have found in the context of the Rindler-Ishak method. Observe that, the contribution of the cosmological constant to the deflection angle, $ \delta\phi_{ \Lambda } $, is divergent when the parameters $ M, Q^2 $ and $\eta'$ goes to zero due to the last term in \eqref{Lambdacontrpsi}, see also \eqref{drpiover2}. To overcome this, we will constraint the domain of the deflection angle by the following setup. Now, in order the deflection to occur from the RN-dS-monopole black hole, the incoming photon orbits should be outside of the photon sphere of the black hole; otherwise, it cannot reach to the observer measuring the deflection angle. The location of the photon sphere can be calculated from (\ref{EffectivePotential})  by taking $V'_{eff}=0$, which yields
\begin{equation}\label{PhotonSphere}
r_{ps} = \frac{1}{2 b^2} \left[ 3 M+ \sqrt{9 M^2-8 b^2 Q^2} \right], \quad b^2 = 1 - 8 \pi \eta^2
\end{equation}
where $r_{ps}$ is the radius of the photon sphere of RN-dS-monopole black hole spacetime, which is independent of $\Lambda$. The effect of the monopole is to increase the radius of the photon sphere. To make a light deflection observation, an incoming photon should not be captured by the black hole. For a photon coming from far away not to be captured by this black hole, from Eq. (\ref{EnergyEquation}), it must satisfy the following inequality 
 \begin{equation}\label{Rbound}
 R^2\ge \frac{27 M^4+8\, b^4\, Q^4-36 b^2\, M^2\, Q^2+(9 M^2-8 b^2  Q^2 )M\sqrt{9 M^2-8 b^2 Q^2}}{2 b^6 M^2-2 b^8 Q^2}:=R_c^2.
\end{equation}
This result gives a lower bound on the impact parameter. For $Q=0$, this inequality reduces to   $R\ge 3\sqrt{3}M/b^3$ and for $b=1$, this inequality reduces to well known result \cite{Misner:1974qy}, $R\ge 3\sqrt{3}M$  for Schwarzschild(-de Sitter) spacetime. The effect of the monopole is to increase the lower bound on the impact parameter. 

We have derived the orbit of photons for this spacetime in Eq. (\ref{secondordersolution}). For a photon following this orbit, the minimum distance of approach of light rays to the black hole, $r_{min}$, given by Eq. \eqref{r:min}, should be outside the photon sphere. Hence, we have the following bound
\begin{equation}\label{lower:bond}
	r_{min}>r_{ps};	
\end{equation}
otherwise, the photon will be captured by the black hole. However, we may not need to evaluate this inequality, since if the impact parameter $R$ already satisfies inequality (\ref{Rbound}), then the incoming light rays would not be captured. Note also that in order for a light deflection observation to take place, the impact parameter must be far greater than $R_c$, namely $R\gg R_c$ \cite{Misner:1974qy} or in other words, we should have $r_{min}\gg r_{ps}$. These results valid for both RN-monopole and RN-dS-monopole spacetimes. Now let us consider the asymptotic nature of the RN-dS-monopole black hole. Near the cosmological horizon, the effect of the mass and charge of the black hole can be ignored; hence, the leading terms in $\Delta_r$ are $\Delta \sim b^2 r^2-\Lambda\, r^4/3$ as $r\gg r_+$. Hence,  the spatial equatorial geometry near the cosmological horizon  can be described by the following line element for $\theta = \pi/2 $, as 
\begin{equation}\label{upper:bond}
	dl^2 = \left( b^2 - \frac{ \Lambda r^2 }{3} \right) dr^2 + r^2 d\phi^2.
\end{equation}
The location of the cosmological horizon can be approximately obtained from this result, since the effects of the mass and charge can be ignored near it, as 
\begin{equation}
	r_c \approx \sqrt{\frac{3 \, b^2}{\Lambda}},
\end{equation}
here, $r_c$ represents the cosmological horizon of the spacetime and it is the upper bound to our radial coordinate. That is, the light source, the deflecting black hole and the observer should be inside the cosmological horizon. Since the monopole term is topological, it affects, namely decreases, the location of the cosmological horizon of dS spacetime.  With these result, we see that to observe light deflection, the orbit of the photon should satisfy the following inequality:
\begin{equation}\label{constraint}
	r_{ps} < r < r_c,
\end{equation}

where $r=1/u$ denotes  the orbit of photon given by (\ref{secondordersolution}) and notice that $r_{min}\gg r_{ps}$ is still applied. Hence by Eq. \eqref{constraint}, we have restricted the observational domain of the deflection angle. Note that for large values of $r$, if we replace  in inequality  (\ref{constraint}), $r$ with its asymptotic values as $\theta\rightarrow \pi/2$, given in Eq. (\ref{rpiover2}), then we can obtain the following limit where the parameters should obey  for the light deflection to be present: 
\begin{equation}
r\simeq \frac{ R}{2\varepsilon}
\left(    
		1 - 2 \eta' + 4 \nu
	+ \frac{ 15 \pi  \eta'\nu }{32 \varepsilon }
	- \frac{ \pi  \eta' }{ 8 \varepsilon }
	+ \frac{ 3 \pi \nu }{ 16 \varepsilon }
	- \frac{ 3 \pi  \eta'^2}{ 32 \varepsilon }
	- \frac{ 15 \pi  \varepsilon }{16}
	- \frac{ 105 \pi  \nu ^2}{256 \varepsilon }
\right)<\sqrt{\frac{3 \, b^2}{\Lambda}}
\end{equation}

 \begin{figure}[!ht]
 	\begin{centering}
 		\includegraphics[draft=false,width=12cm]
 						{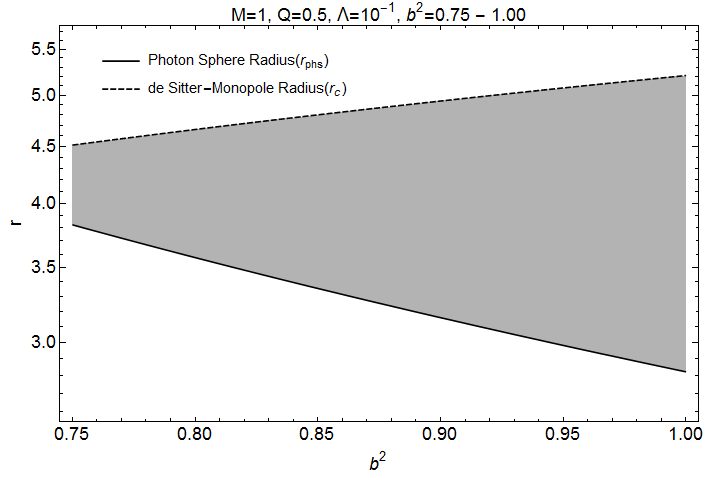}
 		\caption{
 				 Graph showing the change of the observational 
 				 domain (shaded area) of bending of light from a 
 				 RN-dS-monopole black hole, in the interval of 
 				 $b^2= 0.75-1.00$ with $M=1, Q=0.5$ and $
 				 \Lambda=10^{-1}$. For small values of $b^2$(large 
 				 values of $\eta'$), observable domain gets 
 				 smaller. This is since the coupling of the monopole term, 
 				 $b^2$, increases the value of photon sphere 
 				 radius, $r_{phs}$, and decreases the radius of 
 				 cosmological horizon, $r_c$, of the RN-dS-monopole 
 				 system. Note that for graphical purposes, we 
 				 choose an unrealistically big value for the 
 				 cosmological constant compared to its observational 
 				 value.
				}
 		\label{obs:domain}
 	\end{centering}
 \end{figure}
 
From this inequality, we can derive bounds on the parameters of a black hole and impact parameter to ensure that the incoming photons are inside the cosmological horizon such that the deflection angle is not divergent. For example, we can find a bound on the mass and impact parameter  in the linear order for Kottler spacetime,  by setting $b=Q=0,$  in Eq. \eqref{rpiover2}, as:   
\begin{equation}
	\frac{2M}{R^2}>\sqrt{\frac{\Lambda}{3}}.
\end{equation}
This bound gives a lower limit for mass for a given impact parameter, which is irrelevant for astronomical sources.

When we analyze the effect of the monopole term on the observational domain of validity, we see that it increases the photon sphere radius or lower bound and decreases the radius of the cosmological horizon or upper bound; hence, it decreases the domain of validity for the light scattering. In order to see this effect more clearly, we present a graph in Fig. \ref{obs:domain}. In this graph, we have used  a very unrealistic and large value for the cosmological constant to visualize the behavior better,  but this behavior persists for any positive values for $\Lambda$.  Note that, for RN-monopole black holes, there is no such bound because the spacetime is locally asymptotically flat with a deficit solid angle.

%%%%%%%%%%%%%%%%%%%%%%%%%%%%%%%%%%%%%%%%%%%%%%%%%%%%%%%%%%%%%%%%%%%%%%%%%%%%%%%%%%%%%%%%
 
\section{Conclusion}

In this article, we have presented the deflection angle of a photon from both RN-monopole and RN-dS-monopole spacetimes. We have used the perturbative method up to the second-order to solve the modified Binet equation. Using this solution and also using the well-known  Rindler-Ishak method, which helped us to calculate the effects of asymptotically non-flat geometry on the deflection angle, we have successfully obtained the deflection angle in terms of the black hole and monopole terms up to second-order for RN-monopole spacetime. We have seen that the monopole terms act on the deflection angle in an enhancing way.

 Moreover, to obtain the effect of the cosmological constant regarding the deflection angle, we have again exploited the Rindler-Ishak method for the RN-dS-monopole black hole. We see that, in the weak field limit, the interaction term between the gravitational monopole and the cosmological constant effects the bending angle. In both of the cases, our results agree with previous results using the same methods with some corrections we have explained in Sect. 4. These results may be used to constrain the observational bound on the value of the monopole coupling term using the deflection angle of known sources. We have also discussed the domain of validity of these results and see that the monopole term slightly decreases this domain.  A possible next step is to generalize these results to axially symmetric rotating black holes involving a monopole term.

%%%%%%%%%%%%%%%%%%%%%%%%%%%%%%%%%%%%%%%%%%%%%%%%%%%%%%%%%%%%%%%%%%%%%%%%%%%%%%%%%%%%%%%%

\section{Acknowledgements}
M. H. S. and O. D. are supported by Marmara University Scientific Research Projects Committee (Project No: FEN-C-YLP-150218-0055).

% Bibliography
%-----------------------------------------------------------------
\bibliographystyle{ieeetr}
\bibliography{paper}
\end{document}